\documentclass{article}

\usepackage{arxiv}

\usepackage[utf8]{inputenc} 
\usepackage[T1]{fontenc}    
\usepackage{hyperref}       
\usepackage{url}            
\usepackage{booktabs}       
\usepackage{amsfonts}       
\usepackage{nicefrac}       
\usepackage{microtype}      
\usepackage{lipsum}		
\usepackage{graphicx}
\usepackage{natbib}
\usepackage{doi}

\title{RealityDrop: A Multimodal Mixed Reality Framework to Manipulate Virtual Content between Cross-system Displays}

\author{ \href{https://orcid.org/0000-0001-5673-5899}{\includegraphics[scale=0.06]{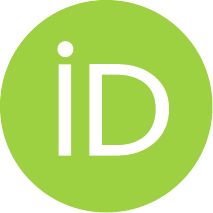}\hspace{1mm}Jeremy McDade}\thanks{} \\
	University of South Australia\\
	\texttt{jeremy.mcdade@unisa.edu.au} \\
	\And
	\href{https://orcid.org/0000-0002-0418-3217}{\includegraphics[scale=0.06]{orcid.pdf}\hspace{1mm}Allison Jing} \\
	University of South Australia\\
	\texttt{allison.jing@mymail.unisa.edu.au} \\
 	\And
	\href{https://orcid.org/0000-0003-2536-3011}{\includegraphics[scale=0.06]{orcid.pdf}\hspace{1mm}Andrew Cunningham} \\
	University of South Australia\\
	\texttt{andrew.cunningham@unisa.edu.au} \\
}

\renewcommand{\shorttitle}{\textit{arXiv} Template}

\hypersetup{
pdftitle={A template for the arxiv style},
pdfsubject={q-bio.NC, q-bio.QM},
pdfauthor={Jeremy McDade},
pdfkeywords={Mixed Reality, Interaction Techniques, Display Technologies},
}

\begin{document}
\maketitle

\begin{abstract}
In this poster, we present RealityDrop, a novel multimodal framework that uses Mixed Reality (MR) technology to manipulate, display, and transfer virtual content across different display systems. Employing MR as the centre of control, RealityDrop affords concise information dissemination among diverse collaborators, through varied representations that best fit each display system's unique features using `superhuman' gaze and gesture interactions. Three multimodal interaction techniques, a customised content interpreter, and two cross-system interfaces are incorporated for fluent content manipulation and presentation. 
\end{abstract}

\keywords{Mixed Reality \and Interaction Techniques \and Display Technologies}

\section{Introduction}
MR-based display systems are introduced each year to improve collaboration in a shared workspace, enabling a vast opportunity for 3D virtual content to be represented based on different levels of control, details and spatial features. It provides the affordance for a team of diverse experts to quickly navigate, present, and discuss information virtualised in the MR task space freely \citep{McDade2022ISMAR}. Conventional 3D interfaces (desktop, mobile or tablets) offer rather simple manipulations, such as changing the sizes (pinch and drag to enlarge or shrink) or rotating the angles of the virtual content among collaborators. The interaction techniques and display angles may not be entirely customized to best accommodate the intrinsic features of information an expert tries to convey \citep{Jing2019}. Moreover, immediate content transfer often seems to require much physical effort across different display systems. For example, we often see a USB chord or external storage used to transfer cross-system information, although researchers have attempted to introduce mouse or stylus-based click-and-drop techniques to relieve such effort \citep{Rekimoto99}. 

\begin{figure}
\centering
  \includegraphics[width=\textwidth]{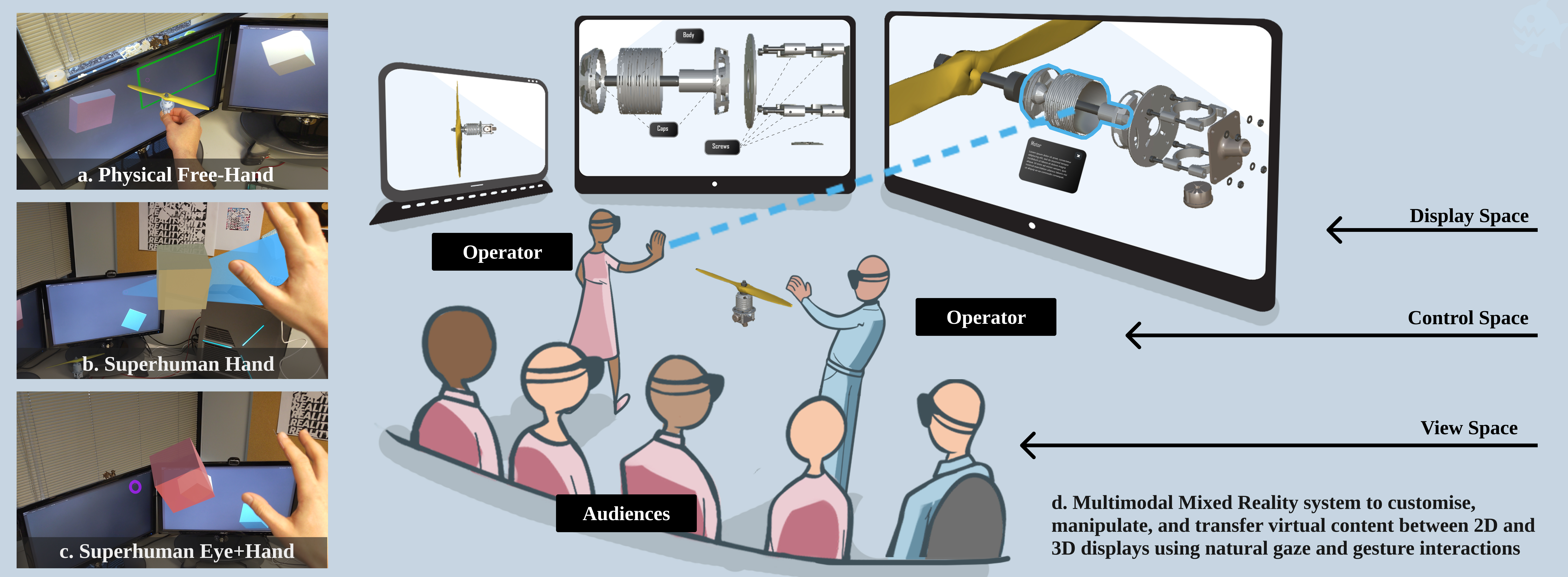}
  \caption{System overview: three interaction techniques (a-c) and a user scenario (d)}
  \label{fig:teaser}
\end{figure}

As a result, we introduce \textbf{RealityDrop\footnote{\url{https://youtu.be/WaqZ5Sdt23A?feature=shared}}} (\autoref{fig:teaser}), a novel multimodal MR framework to customise, present and share virtual content between 2D and 3D displays using gaze and gesture interactions. Our system demonstrates a conceptualized idea to integrate multiple displays using MR as an interactive form of control (control space). It disseminates information based on its volume, variety, spatial and temporal characteristics across domain experts (in view space), to transfer and amplify the presentation on a range of multi-platform display technologies (display space). \textbf{RealityDrop} offers three core concepts: (1) \textit{Three multimodal interaction techniques}, including the basic free-hand drag-and-drop, superhuman hand gestures, and superhuman gaze and hand. Users can retrieve and place content directly to the display space without physically walking up to it. (2) \textit{A customised content interpreter} defines how content should be represented via virtual displays (Augmented Reality, Virtual Reality, and MR) or display devices (conventional 2D display or mobile display). An Operator who manipulates and navigates the virtual content in the MR control space decides what feature best accommodates which display technology according to their expertise in a team. (3) \textit{A cross-system transfer interface} that features two styles (basic and distinct) to indicate how information can be disseminated and amplified to team members (with different expertise) in the view space, allowing for an easy and consistent cross-system user experience.

\section{System Overview}
In (\autoref{fig:teaser}), content can be pre-defined to utilise different display specifications to accommodate the levels of details \textit{operators} (the user who performs interactions in control space) plan to present to the \textit{audiences} (the users in view space) across systems (in display space). Those details include viewing angle, distance, orientation, internal component, etc. For example, in \autoref{fig:teaser}:d, an operator presents an assembled computer-aided design (CAD) model using superhuman hand gestures in multiple displays, where each component of the model is disseminated and displayed differently. 

\subsection{Multimodal Interaction Techniques}
RealityDrop includes three interaction techniques: physical free-hand drag-and-drop, superhuman hand, and superhuman gaze+hand, to control the placement and retrieval of content across display devices. Placement refers to the action of transferring content from the control space to the display space, and conversely, retrieval refers to the opposite.

\textbf{Physical Free-hand Drag-and-Drop}
This baseline technique requires the operator to physically grab content with their hand to place it from the control space into the display space (\autoref{fig:teaser}: a).

\textbf{Superhuman Hand}
This technique uses in-air natural hand gestures to manipulate content from a stationary location in the control space. When interacting with the content, the operator raises their hand with palm forward, like a superhuman absorbing motion, towards a specific display. A ray will be highlighted from the centre of the operator's palm to select and `suck' the content (\autoref{fig:teaser}: b). The content will then `fly' to the operator's hand in the MR control space from the display space. This technique requires no physical movement to a specific display and uses lines connecting the user's palm to the chosen display (palm raycast) as directional cues to help with placement accuracy.

\textbf{Superhuman Gaze+Hand}
Similarly, this technique selects the content using the user's eye gaze. The gaze cursor of the operator's current focus replaces the palm ray to indicate selected content, and the same superhuman gestures are then used to perform the corresponding between-space `absorbing' actions (\autoref{fig:teaser}: c). This technique eases the operator's effort from raising hands to gazing at the content to select.

\subsection{Customised Content Interpreter}
In RealityDrop, virtual content is defined as three-dimensional objects that can be transferred from a range of display scales (wall-size projectors and table-size touchscreens to smartphone and laptop screens). The spectrum of content encompasses entities as simple as cubes to more intricate constructs like data visualizations or computer-aided design (CAD) models. Two spatial concepts are used to define the current positional state of the content. The \textit{control space} is where the operator manipulates and controls the arrangement of content in MR, while the \textit{display space} denotes a specific display location where content is presented. For example, the operator directs a (or part of a) model to an engineer's display, and the interpreter chooses to arrange that content to best suit the engineer's needs and understanding.


\subsection{Cross-system Transfer Interface}
Two cross-system transfer interfaces, basic and distinct, are incorporated to add visual guidance and effects to the transfer process.


\textbf{Basic Transfer Interface}
The transfer process translates the content's position and orientation between the operator's release point and display locations to ensure consistent behaviours during transfer. The basic interface allows for simple adaptations by changing colour, scale and opacity between the control and display spaces. For example, by moving an object to a larger display, the object could increase its scale but keep the colour to fit the dimensions of that display correctly. Meanwhile, the colour and orientation of the last interaction should remain from one space to another.

\textbf{Distinct Transfer Interface}
This interface allows effects to be customised to a specific type of pre-defined content. For example, when operators placed content into different displays, a CAD model could transition from an assembled overview to multiple exploded views with component labels and details (see three displays in \autoref{fig:teaser}: d). Therein, this reinforces the audiences of that display's ability to understand the content.

\subsection{Software and Hardware}
RealityDrop was created in Unity 2021.3.15f1 and built for Microsoft Hololens 2 and Windows. Any MR device using the Mixed Reality Toolkit (MRTK) or any display supporting Unity can be utilized by RealityDrop. The interaction component uses controller systems from the MRTK, and networking integration uses the Photon Unity Networking 2 (PUN2) plugin.

\subsection{Use Case}
Dissemination of meaningful information with a diverse range of experts using customized visualizations presents an exciting use case for RealityDrop. To illustrate, mechanical engineers, designers, and project managers can all perceive identical 3D content in distinct ways. It is imperative to show each strength of the individual. The operator can use the custom content interpreter to manipulate MR virtual parts of a 3D CAD model, and then present them to the audience in a manner that aligns with their expertise. Therefore, engineers view a mechanical schematic, designers focus on colours and text, while project managers see an overarching summary such as \autoref{fig:teaser} d. Our video demo at \url{https://youtu.be/WaqZ5Sdt23A?feature=shared} presents those use cases.

\section{Conclusion \& Future Work}
This poster proposes a multimodal MR framework to customise, manipulate and transfer virtual content between 2D and 3D display space. The operator interacts and manipulates virtual content in the MR control space, best amplifying and transferring content to the corresponding display using `superhuman' gaze and hand gestures, among diverse experts to collaborate. In the future, we plan to conduct a full user study to compare how interaction techniques (baseline, gestures, gaze+gestures) affect displayed information (customised and non-customised) among groups of domain experts.

\bibliographystyle{unsrtnat}


\end{document}